\documentstyle[aps,epsf,epsfig]{revtex}
\begin{document}

\def\be{\begin{equation}}
\def\ee{\end{equation}}
\def\ba{\begin{eqnarray}}
\def\ea{\end{eqnarray}}
\def\la{~\mbox{\raisebox{-.6ex}{$\stackrel{<}{\sim}$}}~}
\def\ga{~\mbox{\raisebox{-.6ex}{$\stackrel{>}{\sim}$}}~}
\def\bq{\begin{quote}}
\def\eq{\end{quote}}
\def\PL{{ \it Phys. Lett.} }
\def\PRL{{\it Phys. Rev. Lett.} }
\def\NP{{\it Nucl. Phys.} }
\def\PR{{\it Phys. Rev.} }
\def\MPL{{\it Mod. Phys. Lett.} }
\def\IJMP{{\it Int. J. Mod .Phys.} }
\font\tinynk=cmr6 at 10truept
\newcommand{\bfx}{{\bf x}}
\newcommand{\bfy}{{\bf y}}
\newcommand{\bfr}{{\bf r}}
\newcommand{\bfk}{{\bf k}}
\newcommand{\bkp}{{\bf k'}}
\newcommand{\order}{{\cal O}}
\newcommand{\beq}{\begin{equation}}
\newcommand{\eeq}{\end{equation}}
\newcommand{\beqa}{\begin{eqnarray}}
\newcommand{\eeqa}{\end{eqnarray}}
\newcommand{\mpl}{M_{Pl}}
\newcommand{\lmk}{\left(}
\newcommand{\rmk}{\right)}
\newcommand{\lkk}{\left[}
\newcommand{\rkk}{\right]}
\newcommand{\lnk}{\left\{}
\newcommand{\rnk}{\right\}}
\newcommand{\Rbar}{\bar{R}}
\newcommand{\gbar}{\bar{g}}
\def\la{~\mbox{\raisebox{-.6ex}{$\stackrel{<}{\sim}$}}~}
\def\ga{~\mbox{\raisebox{-.6ex}{$\stackrel{>}{\sim}$}}~}

\def\ltap{\ \raise.3ex\hbox{$<$\kern-.75em\lower1ex\hbox{$\sim$}}\ }
\def\gtap{\ \raise.3ex\hbox{$>$\kern-.75em\lower1ex\hbox{$\sim$}}\ }
\def\gl{\ \raise.5ex\hbox{$>$}\kern-.8em\lower.5ex\hbox{$<$}\ }
\def\roughly#1{\raise.3ex\hbox{$#1$\kern-.75em\lower1ex\hbox{$\sim$}}}

\twocolumn[\hsize\textwidth\columnwidth\hsize\csname
@twocolumnfalse\endcsname
\preprint{hep-th/0211221\\ November 2002}
\date{\today}
\title{Holographic Limitations of the Effective Field Theory of Inflation}
\author{Andreas Albrecht, Nemanja Kaloper and Yong-Seon Song}
\vskip.5cm
\address{{\em Department of Physics, University of California, Davis, CA 95616}\\}
\maketitle
\begin{abstract}
Holographic considerations may provide a glimpse of quantum
gravity beyond what is currently accessible by other means. Here
we apply holography to inflationary cosmology. We argue that the
appropriate holographic bound on the total entropy in the
inflationary perturbations is given by the {\it difference} of the
area of the apparent horizon in the slow-roll approximation and
its exact value near the end of inflation.  This implies that the
effective field theory of the inflaton weakly coupled to gravity
requires modifications below the Planck scale.
For a conventional model of inflation this scale is of order of the GUT scale,
$\Lambda_{UV} \la 10^{16} GeV$, but could be considerably lower.
Signatures of such new physics could show up in the CMB.
\end{abstract}
\pacs{PACS: 98.80.Cq \hfill  hep-th/0211221 }
\vskip-1pc]

We live in a big, old, but nearly wrinkle-free universe. Inflation
is currently the best paradigm which could explain how it got its
face-lift \cite{infl}. However, a truly fundamental understanding
of inflation is yet to emerge from yet to be discovered complete theory of quantum
gravity.  We are therefore far from building cosmology directly from a
complete quantum gravity theory (see, e.g.
\cite{old}-\cite{Witten}). Nevertheless, there have been
remarkable developments in black hole physics which led to the
discovery of general principles, originating from black hole
thermodynamics, that appear to be universal and transcend any
given specific approach to quantum gravity. 't Hooft \cite{thooft}
and Susskind \cite{lenny1} have named this set of rules the
holographic principle, showing that for black holes, it is {\it de
facto} the solution of the black hole information problem. It
states that a black hole has far fewer degrees of freedom than
expected, with its total entropy given by the quarter of the
horizon area in Planck units, as per Bekenstein-Hawking formula
\cite{bek,hawk}.  It has since been shown that generic field
theories coupled to gravity overcount the degrees of freedom,
by deducing ``holographic area bounds'' which limit the number of
physical degrees of freedom. In this Letter we explore the
implications of these ideas for inflation.

Fischler and Susskind have made the first step towards incorporating
the holographic principle in cosmology \cite{fs}, suggesting that
the entropy which crosses the lightlike boundary of an observable
region of the universe, the particle horizon, should not exceed
the horizon area in Planck units. Subsequent to
\cite{fs,otherholos}, Bousso has offered a covariant version of
the holographic bound \cite{bousso}, outlining a prescription for
choosing the holographic ``screens". The idea is to take any
lightlike surface ending on either a caustic or the apparent
horizon, and limit the entropy which crosses it by a 1/4 of the
maximal area along the light sheet in Planck units. This seems to
produce desirable results for matter with generic properties
\cite{fmw}. The implications of holography for cosmology have been
explored in many works (e.g. \cite{Banks}-\cite{Witten},
\cite{dks}, \cite{tomwilly}, \cite{hogan}).

Here we argue that the holographic principle imposes a limit on
the validity of the description of inflation, and specifically
inflationary fluctuations, via an effective field theory weakly
coupled to gravity in the causal patch (EFTwg). The crux of our
argument is that the entropy deposited into the metric
fluctuations generated by the inflaton could easily exceed the
holographic bounds, if the EFTwg is valid to arbitrarily short
distances. This cannot happen in a truly holographic theory. An
example is a field theory in a black hole background, where the
entropy, defined as the logarithm of the partition function, in a
causal region of a black hole spacetime, is badly divergent
\cite{thooftbw}. Only after renormalizing it does one recover the
Bekenstein-Hawking area law \cite{thooftbw}. This corresponds to
cutting off the modes in the UV, residing close to the horizon. We
argue that in inflation this new physics encoded in the UV cutoff
affects the inflationary fluctuations, which are classical scalar
modes with their wavelengths greater than the instantaneous Hubble
length. These modes start out as short distance excitations inside
the Hubble horizon before they are expelled out by the
superluminal expansion, and hence are sensitive to the physics
represented by the UV cutoff.

In the absence of a precise formulation we can crudely
approximate the holographic redundancies by an energy cutoff
$\Lambda_{UV}$ on EFTwg, which remains sensitive to the dynamics.
This cutoff is a reflection of the expected strong gravitational interactions
close to the horizon.
We find that it is bounded from above, and the bound is inversely
proportional to the third root of the number of efolds. This
implies that the number of efolds which can be described by EFTwg
in the causal patch is limited. In the commonly quoted example of inflation occurring
when $H \la 10^{14} GeV$, and lasting about $\sim 100$ efolds, the
cutoff is $\Lambda _{UV} \la 10^{16} GeV \sim M_{GUT}$, but it could
be lower depending on the model of inflation. A similar
scale also appears as the upper bound on the scale of new physics
which might imprint an observable signature in the CMB
as the ${\cal O}(H^2/\Lambda^2_{UV})$ corrections on top of the leading
order fluctuations \cite{kkls}. We submit that these signatures
could be a reflection of holography, although a deeper knowledge of the
underlying physics is required to determine their detailed form.

We start with the picture of pure de Sitter space as a ``hot tin
can''\cite{dks}, with an observer in the causal patch (the
interior of the ``can'') surrounded by a hot horizon (the
``walls''). The physical degrees of freedom, accessible to the
observer, are in thermal equilibrium with the walls, at the
Gibbons-Hawking temperature. In the case of nonzero, constant slow
roll parameters, such as in a quintessence-dominated universe, the
picture of \cite{accel} suggested that the ensuing causal region
can be viewed as a hot tin can with moving walls, so that the
temperature of the gas in equilibrium inside the cavity decreases
due to the expansion of the can. An inflating region in the slow
roll regime closely resembles de Sitter space, with a key
difference: unlike the perfect de Sitter space, where the apparent
horizon and the event horizon coincide and so there is no
``exterior" of the can, during inflation modes in the IR are
expelled from the apparent horizon. These modes are real physical
degrees of freedom, and may or may not ever thermalize depending
on the future evolution of the universe. We will divide the total
instantaneous entropy in the inflating patch into the dominant
equilibrium contribution, approximated by the horizon area of the
would-be pure de Sitter space, and a small non-equilibrium
contribution, where the inflaton fluctuations count in. We will
argue that this is analogous with the familiar case of a black
hole in de Sitter space. In our picture inflation is a hot,
expanding, but porous tin can, slowly leaking entropy into a cold
reservoir via the expelled modes.

Assuming the standard theory of inflationary fluctuations below the
cutoff $\Lambda_{UV}$, we estimate the entropy-per-mode by
counting the mode's quantum excitation level
\cite{grishchuk,bmp}. We argue that this entropy is bounded by the
{\it difference} of the areas of true apparent horizon during
inflation, and its slow-roll value. This goes beyond the covariant
holographic bounds \cite{bousso} in a crucial way, because it
allows us to discriminate between the entropy of modes in
equilibrium and the entropy in density fluctuations which froze
out. Our bounds constrain $\Lambda_{UV}$ to physically
interesting values, which depend on the inflation model, but in
general are significantly below $M_p$.

We will work with the flat $4D$ FRW metric and ignore
any non-inflaton matter sources (both good approximations during
inflation). The background obeys
\ba
\label{eoms}
3H^2 &=& \frac{{\dot \phi}^2}{2M_p^2} + \frac{V(\phi)}{M_p^2} \, ,
~~~~~ \ddot \phi + 3H \dot \phi + \frac{\partial V(\phi)}{\partial \phi} = 0 \, ,
\ea
where the overdot is the time derivative, $H = \dot a/a$ is the
Hubble parameter, $a$ is the FRW scale factor, $\phi$
is the inflaton, and $V(\phi)$ is the inflaton potential. The
Planck mass is defined by $M_p = 1/\sqrt{8\pi G_N}$ since
$c=\hbar=1$.

Inflation occurs when the potential energy of the scalar dominates
in (\ref{eoms}). The dynamics is well approximated by the
slow-roll regime, formally defined by the standard conditions
$|\epsilon|, |\delta| \ll 1$ where the slow-roll parameters are
$\epsilon = {3 \dot \phi^2 \over 2 V} $,
$\delta = {\ddot \phi \over H \dot \phi}$.
The slow-roll approximation then consists of dropping the terms of
order ${\cal O}(\epsilon,\delta)$. Then
(\ref{eoms}) reduce to
\be
3H^2 = {V \over M^2_p}   \, , ~~~~~~~~~~
3H \dot \phi + {\partial {V} \over \partial \phi} = 0 \, .
\label{slowreqs}
\ee
We stress here that in the slow-roll approximation the potential,
and thus the radius of the apparent horizon (i.e. the Hubble horizon),
defined as the hypersurface at $l_{AH} = H^{-1} = M_p/\sqrt{3V}$ from
the origin, is essentially constant during each Hubble time.
Although $V$ is implicity
time-dependent, as seen from Eq.
(\ref{slowreqs}), it changes during a single efold by
at most $\Delta V \la |\frac{V' \Delta \phi}{V}| \simeq
\frac{\epsilon}{2} V$. Because one {\it defines} the slow-roll
by dropping ${\cal O}(\epsilon,\delta)$ terms, the
solutions are only reliable to this order, and so the radius of
the apparent horizon is constant with the
error of only ${\cal O}(\epsilon)$. This is crucial for our
arguments to follow.

\begin{figure}[htb!]
\hspace{2.8truecm}
\epsfysize=1.8truein
\epsfbox{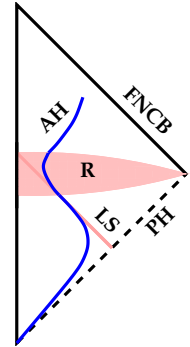}
\caption[]%
{\small\sl Penrose diagram of an inflationary spacetime. }
\end{figure}
\noindent

The causal structure of the
inflating region is given in Fig. 1 \cite{kkls}.
The future null line is the future null causal boundary (FNCB),
which could be either a null infinity, or an event horizon if the
universe ends up inflating forever. The past null line is the
particle horizon (PH), measuring how far the light travelled since the
birth of the universe (taken as $t=0$), given by $L_H(t) = a(t)
\int^t_0 \frac{dt'}{a(t')}$. The shaded area (R) represents reheating,
with inflation occurring before it, and the radiation/matter era
after it. The apparent, i.e. Hubble, horizon (AH) is the bold solid
curve, during inflation almost null and directed first outward,
turning inward later. This is because $H \approx$ const. during
inflation, and reflects the thermal nature of the quantum state
during inflation \cite{kkls}, an approximation valid over the time
scales of order the Hubble time as argued above. According to the
complementarity principle, all active experiences of an observer
can be fully described by the physics in the causal patch, or the
tin can \cite{Banks,dks,dls,accel}. As we are only interested in
the inflationary fluctuations pushed in between AH and FNCB,
we will ignore the exterior of the causal
patch as immaterial.
LS is a light-sheet defining a holographic screen.

The out-of-equilibrium modes originate from within the apparent horizon,
and are expelled out of it by inflationary expansion. They
remain frozen until they cross back
into the apparent horizon after inflation, where they grow
due to the gravitational instability. The entropy per mode is
found by considering the fluctuations as squeezed states of the
inflaton \cite{grishchuk,bmp}. One should sum up the entropy in all modes
which were expelled out of the apparent horizon during inflation,
with momenta determined by the horizon crossing matching condition
and the full range of the Hubble scale during inflation
where EFTwg is valid, $H(t) \in
[H, \Lambda_{UV}]$. The key point is
that the result is UV-dominated, since the
inflationary fluctuations are nearly scale-invariant, with
$P_\Phi(p) \sim k^{n-1}$ where $n = 1 + {\cal O}(\epsilon,\delta)$
is close to unity as $|\epsilon|,|\delta| \ll 1$, and so we can
{\it i)} approximate the entropy per mode by an (almost) constant
value, of order unity and {\it ii)} assume that the modes are
pushed out of the horizon at a rate independent of their
wavelength. The entropy deposited in them will only depend on
the number of modes pushed out of the horizon. The modes which left
the horizon at the beginning will be
the most numerous, since at that time the Hubble length was the
shortest possible, and they will dominate the entropy. Because
their wavelength was of order of the Hubble length in the beginning of the
EFTwg description of inflation, $\lambda_{UV} \simeq \Lambda_{UV}^{-1}$, their number
per horizon volume at the end of inflation is of order of the number
of the initial Hubble domains inside the thermal patch at the end
of inflation, $n \simeq 4\pi (\lambda_{thermal}/\lambda_{UV})^3
\simeq 4\pi (\Lambda_{UV}/H)^3$. Thus, the entropy in the expelled
modes is
\be
S \simeq 4\pi \frac{ \Lambda^3_{UV}}{H^3} \, .
\label{entrcut}
\ee
The calculations of \cite{grishchuk,bmp} bear out this estimate.
Eq. (\ref{entrcut}) presumes that fluctuations leave
the apparent horizon, which is only possible if it is timelike,
so that the slow roll parameters are small but not exactly zero. To
take the limit $\epsilon,\delta \rightarrow 0$
one must take $\Lambda_{UV} \rightarrow H$
and keep the subleading corrections to (\ref{entrcut}), where we
expect $S \rightarrow 0$.

The holographic principle gives us another way of estimating the
entropy in the inflating patch. A rule for
bounding the entropy is to pick a light-sheet crossed by the
entropy, and limit the entropy by the maximal area along the
light-sheet \cite{fs,bousso}. A suitable light-sheet is given by
the faint solid null line (LS) in Fig. 1, which is the
backward light cone of an observer near the end of inflation,
with maximal area ${\cal A} \simeq M^2_p/H^2$. We chose it so
that it maximizes the bounds, as it intersects the apparent
horizon as close as possible to the end of inflation, where the
slow-roll conditions still hold, and so it is crossed by (almost)
all of the modes expelled out of the apparent horizon. However,
the ensuing bounds still are not very strong \cite{fs}, because
they do not discriminate between the modes in equilibrium and those out of it.

We propose here that the bounds could be strengthened by separating these
modes from each other during inflation. To motivate the
procedure we first consider black holes in de Sitter space
\cite{gibbhawk}. They are the most stable objects in a background
de Sitter space: they can be viewed as small, hot, balls of gas in
a cold cavity, which are very slowly melting away by emitting
Hawking radiation. Because this process is very slow, given by the time
$\tau_{BH} \simeq M^3_{BH}/M^4_p$, most of the
energy which slipped into a black hole will not thermalize with de
Sitter background for a very long time. Their entropy is given by their horizon area,
which for black holes smaller than the de Sitter radius is as in
the flat space, $S_{BH} \sim M^2_p {\cal A}_{BH} \sim 4\pi M^2_p
r^2_{BH} \sim (M_{BH}/M_p)^2$. On the other hand, the equilibrium
entropy in the asymptotically de Sitter geometry is approximately
$S_{eq} \sim 4\pi M^4_p/\lambda - {\cal O}(1) M^2_p \sqrt{{\cal
A}_{dS} {\cal A}_{BH}}$. Thus the entropy in the black hole, in
modes out of equilibrium with de Sitter cavity, is smaller than
the {\it difference} between the entropy of the modes in
equilibrium and the total entropy in pure de Sitter space with the
same value of the cosmological constant:
\be
\label{difflaw}
S_{BH} \la \sqrt{S_{BH} S_{eq}} \la S_{dS} - S_{eq} \, .
\ee

We will now adopt a similar formula for the case of inflation. If
the slow-roll parameters were exactly zero, the apparent horizon
would have coincided with the event horizon, and nothing would
ever leave it (or freeze out) in a finite observer's time. Thus in
this limit the entropy in the metric fluctuations would vanish
too. Slow time-dependence of $H$ is thus essential for producing the
fluctuations. However, in the slow-roll approximation during the
times $\la H^{-1}$ we have seen above that the { Hubble} horizon is
{ practically} indistinguishable from a {\it true} event horizon,
{ since $\Delta V = 0$ to leading order in $\epsilon$}. Hence the
slow-roll { Hubble} horizon has no ability to distinguish the modes
in and out of equilibrium. Its area in the slow-roll regime
encodes the contribution from all the field theory modes generated
in the Hubble patch. In contrast, the { true} apparent horizon
discriminates between the modes in equilibrium, with wavelengths
shorter than the Hubble length, and the modes pushed out of
equilibrium, which grow greater than the Hubble patch. In fact,
the apparent horizon moves inward toward the center of the
inflating region { relative to the slow roll Hubble horizon},
as does the cosmological horizon in
Schwarzschild-de Sitter space-time, to record that there is fewer
modes in equilibrium. In a manner of speaking, if exact de Sitter
space is a cavity with hot static walls, the inflating patch is a
cavity with hot, moving and porous walls, which lets out some of the IR
degrees of freedom. To extract their entropy, we take the difference of the
slow-roll and exact areas of the apparent horizon,
\be
\label{deltaa}
\Delta {\cal A} = \frac{\epsilon}{3} \, {\cal A} \, .
\ee
To leading order in the slow-roll parameters we can take
${\cal A}$ on the RHS to be the slow roll horizon area, or
equivalently the horizon area of the would-be de Sitter space
found in the limit $\epsilon,\delta \rightarrow 0$, since the
errors are of order ${\cal O}(\epsilon^2)\ll 1$. Then the field
theory result for the entropy cannot exceed this area in Planck
units,
\be
S \la \frac{\epsilon}{3} \, M^2_p {\cal A} \, .
\label{ourbound}
\ee
This bound on the entropy in inflationary
perturbations is stronger than the covariant, all-inclusive,
bound of \cite{bousso} by the power of $\epsilon \ll 1$.
Using (\ref{entrcut}), we rewrite this as the bound on the
UV cutoff $\Lambda_{UV}$:
\be
\label{boundcut}
\Lambda_{UV} \la \Bigl(\frac{\epsilon}{3} \Bigr)^{1/3} \Bigl(H M^2_p\Bigr)^{1/3} \, .
\ee
This is our main result. It tells us that the causal patch
description of inflation by the effective field theory weakly coupled to gravity cannot
be valid down to arbitrarily short distances. Instead, the
holographic redundancies must cut off the entropy that can fit in the
apparent horizon, including the portion in the IR which can get
pushed out of equilibrium. Let us consider a simple explicit
example applying (\ref{boundcut}) to chaotic inflation
\cite{lindechaotic}. A typical chaotic inflation potential is of
the form $\lambda \phi^n/n$ for an arbitrary (integer) power $n$.
The slow roll parameter $\epsilon$ is related to the number
of efolds before the end of inflation
${\cal N}$ by $\epsilon = \frac{n}{4{\cal N}}$, so
we can rewrite (\ref{boundcut}) as
\be
\label{boundchao}
\Lambda_{UV} \la \Bigl(\frac{n}{12 {\cal N}} \Bigr)^{1/3}
\Bigl(H M^2_p\Bigr)^{1/3} \, .
\ee
Thus, with $H \la 10^{14} GeV$ and
${\cal N} \sim 100$, we get
\be
\label{numconst}
\Lambda_{UV} \la 10^{16} GeV \, ,
\ee
so EFTwg receives modifications below the GUT scale.
A longer inflation requires a
lower cutoff. Note that (\ref{boundcut})-(\ref{numconst}) are approximations,
valid as long as $\Lambda_{UV} \gg H$ and $1 \gg \epsilon > 0$, and that
they are a bound on the semiclassical regime of inflation.
It would be interesting to see what is the relation between eternal inflation
and the causal patch description \cite{spc}.

Our conclusion is that the holographic principle limits the
validity of the causal patch EFTwg description of the inflaton, implying
modifications at a scale set by (\ref{boundcut}), below
$M_p$. We do not necessarily view this as a {\it conceptual}
obstruction to inflation. We see our result as a hint that the
inflaton need not be an elementary field, but emerges from the
fundamental theory merely as an order parameter. At energy scales
above the cutoff it would contain too much entropy, and so its
EFTwg description cannot remain valid to such high energies.
Alternatively, it cannot model an arbitrarily long period of
inflation. An interesting consequence of our arguments is related
to the recent field theory analyses of the signatures of high
energy physics in the CMB. There scales similar to our cutoff
arose as the upper bounds on scales of the new physics which might
imprint subleading signatures on the CMB \cite{kkls}.
Because we expect the usual effective field theory to
hold at scales $H \la 10^{14} GeV < \Lambda_{UV}$ (so that we have
an effective field theory model of inflation in the first place),
the effects of the holographic redundancies could be similar to
the signatures of new physics near the cutoff $\Lambda_{UV}$. In
the EFTwg regime the corrections to the leading order results
appear as ${\cal O}(H^2/\Lambda_{UV}^2)$ effects \cite{kkls}. If
the 
new physics is below the GUT scale the
corrections could beat the cosmic variance,
and may be an exciting prospect for the observational
signatures of holography in the sky.

We thank T. Banks, W. Fischler, M. Kaplinghat, M. Kleban, L. Kofman, A. Lawrence,
A. Linde, S. Shenker and L. Susskind for valuable discussions. A.A. and Y-S.S.
are supported by DOE grant DE-FG03-91ER40674.
N.K. was supported by the UCD,
by NSF Grant PHY-9870115 to the Stanford ITP, and
by NSF Grant PHY-99-07949 to the UCSB KITP.



\begin{thebibliography}{99}
\bibitem{spc}
In fact, the bound (\ref{boundchao}) suggests
that in the causal patch EFTwg the vev of the inflaton
must be below the self-reproduction value. It will be interesting
to further explore the implications of this.
\end{thebibliography}
\end{document}